\documentclass{article}
\usepackage{spconf,amsmath,graphicx}

\usepackage{multirow}
\usepackage{array,booktabs,arydshln,xcolor}

\title{Study on the Correlation between Objective Evaluations and Subjective Speech Quality and Intelligibility}
\name{Hsin-Tien Chiang$^1$, Kuo-Hsuan Hung$^2$, Szu-Wei Fu$^3$,  Heng-Cheng Kuo$^1$, Ming-Hsueh Tsai$^{4}$, Yu Tsao$^{1}$}
\address{$^1$Academia Sinica,
  $^2$National Taiwan University,
  $^3$NVIDIA,
  $^4$National Academy for Educational Research}

\copyrightnotice{979-8-3503-0689-7/23/\$31.00~\copyright2023 IEEE}
\begin{document}
%
\maketitle
\begin{abstract}
Subjective tests are the gold standard for evaluating speech quality and intelligibility; however, they are time-consuming and expensive. Thus, objective measures that align with human perceptions are crucial. This study evaluates the correlation between commonly used objective measures and subjective speech quality and intelligibility using a Chinese speech dataset. Moreover, new objective measures are proposed that combine current objective measures using deep learning techniques to predict subjective quality and intelligibility. The proposed deep learning model reduces the amount of training data without significantly affecting prediction performance. We analyzed the deep learning model to understand how objective measures reflect subjective quality and intelligibility. We also explored the impact of including subjective speech quality ratings on speech intelligibility prediction. Our findings offer valuable insights into the relationship between objective measures and human perceptions.
\end{abstract}
\begin{keywords}
Objective measures, subjective listening tests, speech quality, speech intelligibility
\end{keywords}
\section{Introduction}
\label{sec:intro}
Speech quality and intelligibility are crucial in various speech-related applications, such as speech enhancement (SE), teleconferencing, voice conversion and text-to-speech, and hearing aids. As humans are the end-users of these applications, subjective listening tests are considered the most precise and trustworthy way to evaluate speech quality and intelligibility. However, conducting listening tests on a large number of participants is time-consuming and expensive. Therefore, a significant amount of research has been devoted to developing objective measures that can mathematically quantify speech quality and intelligibility. 
\par Objective measures can be divided into intrusive measures, in which quality and intelligibility are estimated by comparing degraded/processed speech with clean references, and non-intrusive measures, in which quality and intelligibility are calculated directly on the degraded/processed speech without a clean reference. Perceptual Evaluation of
Speech Quality (PESQ) \cite{recommendation2001862} and Perceptual Objective Listening Quality Analysis (POLQA) \cite{beerends2013perceptual} are intrusive speech quality measures. Despite being widely used in speech processing research, PESQ and POLQA have been shown to correlate suboptimally with subjective tests \cite{reddy2019scalable}. The short-time objective intelligibility measure (STOI) \cite{taal2011algorithm} and extended STOI (ESTOI) \cite{jensen2016algorithm} are popular intrusive speech intelligibility measures. However, STOI has been reported to provide suboptimal prediction capability for the subjective intelligibility results of the Wiener filtering \cite{yamamoto2017predicting} or deep learning (DL)-based \cite{gelderblom2017subjective} SE systems. Moreover, intrusive measures are less applicable to real-world scenarios because clean signals may not always be available. Compared to intrusive methods, non-intrusive methods such as ITU-T P.563 \cite{malfait2006p}, ANIQUE+ \cite{kim2007anique+}, and speech-to-reverberation modulation ratio (SRMR) \cite{falk2010non} overcome this limitation. 
\par A recent approach to non-intrusive methods directly predicts objective measures by using DL models without the need for a clean signal. These models were trained to predict standard objective measures, such as PESQ and STOI \cite{fu2018quality,jia2020deep,dong2020attention}. Several studies have demonstrated high performance using this approach. However, the ground-truth labels used to train these DL models are not always aligned with human perception. To better align with human perception, researchers have begun to rely on ground truth human labels for model training. DNSMOS \cite{reddy2021dnsmos} and NISQA \cite{mittag2021nisqa} are examples of DL models trained on mean opinion score (MOS) datasets, where DNSMOS focuses on distortions in SE and NISQA focuses on distortions in communication networks. Andersen \emph{et al.} \cite{andersen2018nonintrusive} and Pedersen \emph{et al.} \cite{pedersen2020neural} used convolutional neural network models to predict subjective intelligibility. However, owing to the time-consuming nature of conducting subjective listening tests, collecting large-scale datasets of human labels to train DL-based models is challenging.
\par One potential solution to bridge the gap between objective measures and human perception without relying on large-scale datasets of human labels is to predict human perception of speech quality and intelligibility by leveraging commonly used objective measures. The advantage of this approach is that it is considerably less time-consuming than conducting subjective listening tests. Previous studies have attempted to predict either speech quality or intelligibility using objective measures. Hu \emph{et al.} \cite{hu2007evaluation} proposed composite measures for evaluating SE by linearly combining objective quality measures. Liu \emph{et al.} \cite{liu2006assessment} showed that automatic speech recognition (ASR) and objective quality measures have the potential to estimate intelligibility under noisy conditions. Ma \emph{et al.} \cite{ma2009objective} reported that objective measures originally designed to predict speech quality could reliably predict the intelligibility of noise-suppressed speeches. However, there is limited research that examines both quality and intelligibility criteria and interprets how objective measures reflect subjective quality and intelligibility in practical use.
\par In this study, we first time proposed using DL models that use a combination of off-the-shelf objective measures as inputs to predict subjective quality and intelligibility ratings. We evaluated the correlation between commonly used objective measures and subjective ratings of quality and intelligibility on a Chinese dataset called TMHINT-QI \cite{chen2021inqss}, and then use DL techniques to propose new objective measures composed of all of the used objective measures. We demonstrated that the proposed DL model can achieve strong performance in predicting subjective quality and intelligibility ratings, even when trained on small amounts of training data. This core strength makes the DL model practical for real-world applications because it can still maintain high accuracy without requiring a large amount of training data. Furthermore, we interpreted the proposed DL models to describe the relationship between the objective measures and subjective ratings of speech quality and intelligibility. We also investigated the potential improvements in intelligibility prediction by incorporating subjective quality ratings. Our results can provide valuable insights into the utility and limitations of objective measures in reflecting subjective quality and intelligibility ratings and potentially contribute to bridging the gap between objective measures and human perception.
\par The remainder of this paper is organized as follows. Section \ref{sec:objectivemeasure} describes the objective measures used in our experiments. Section \ref{sec:tmhintqi} details the dataset and presents the correlation analysis. We present our experimental setup and results in Section \ref{sec:results}. Finally, we conclude the paper in Section \ref{sec:conclude}.

\section{Objective measures}
\label{sec:objectivemeasure}


\subsection{Intrusive objective measures}
Six different intrusive objective measures were assessed: PESQ, ITU-T P.835, normalized covariance metric (NCM), STOI, ESTOI, and word error rate (WER). PESQ evaluates speech quality and ranges from -0.5 to 4.5. ITU-T P.835 evaluates speech quality based on three aspects: signal quality (SIG), background noise (BAK), and overall quality (OVRL) \cite{hu2007evaluation}. The NCM assesses the covariance between the envelopes of the clean and degraded/processed speech and provides scores ranging from 0 to 1 \cite{ma2009objective}. The STOI and ESTOI evaluate speech intelligibility and have scores between 0 and 1. Finally, the WER was calculated using Google ASR. 

\subsection{Non-intrusive objective measures}
Two non-intrusive objective measures were also evaluated: DNSMOS P.835 \cite{reddy2022dnsmos} and MOSA-Net \cite{zezario2022deep}. DNSMOS P.835 is a multi-stage self-teaching based model that evaluates speech quality based on three aspects: signal quality (DNSMOS-SIG), background noise (DNSMOS-BAK), and overall quality (DNSMOS-OVRL). MOSA-Net uses time, spectral features, and latent representations from a self-supervised model and was originally trained to predict several objective metrics, but can be adapted for MOS predictions. We adopted the MOS prediction results of the MOSA-Net. 
\par Four quality measures (PESQ, ITU-T P.835, DNSMOS P.835 and MOSA-Net) and four intelligibility measures (NCM, STOI, ESTOI and WER) were used. We obtained several objective measures, such as WER, DNSMOS, and MOSA-Net, by leveraging pre-trained APIs from third-party sources, eliminating the need for additional efforts to acquire these pre-trained models or gather extensive amounts of training data.  

\section{Correlations between objective and subjective assessments}
\label{sec:tmhintqi}
\subsection{Dataset}
\label{sec:dataset}
We conducted the experiments using TMHINT-QI \footnote{TMHINT-QI dataset: http://gofile.me/6PGhz/4U6GWaOtY; TMHINT-QI dataset description: https://github.com/yuwchen/InQSS} \cite{chen2021inqss}, a Chinese corpus containing noisy and enhanced data. To generate noisy data, we corrupted the clean speech from the TMHINT dataset with four types of noise (babble, street, pink, and white) at four different SNR levels (-2, 0, 2, and 5 dB). The noisy data were then enhanced by the minimum mean squared error (MMSE), Karhunen-Loéve transform (KLT), deep denoising-autoencoder (DDAE), fully convolutional network (FCN), and transformer model (denoted as Trans). 

\begin{figure*}[t!]
\begin{minipage}[b]{1\linewidth}
  \centering
  \centerline{\includegraphics[width=16cm]{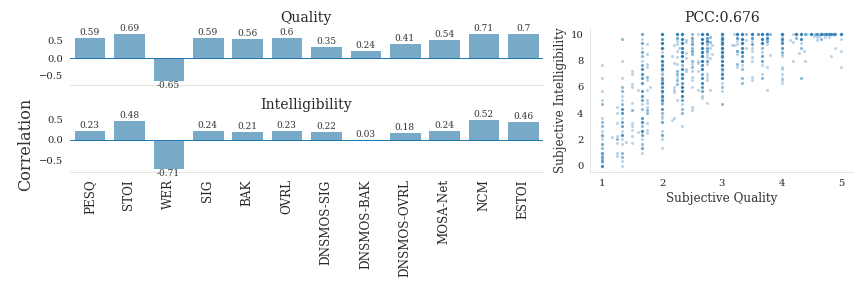}}
\end{minipage}
\caption{Bar plot depicting the Pearson correlation coefficient between subjective and objective measures (left), and scatter plot of subjective quality versus subjective intelligibility (right).}
\label{fig:Corr}
\end{figure*}

\begin{figure}[t!]
\begin{minipage}[b]{1\linewidth}
  \centering
  \centerline{\includegraphics[width=9.5cm]{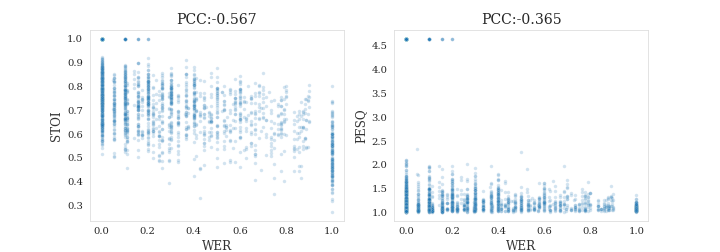}}
\end{minipage}
\caption{Scatter plots depicting the correlation between WER and estimated objective measures.}
\label{fig:DetailCorr}
\end{figure}

\par Human listeners were recruited to evaluate the subjective TMHINT-QI scores. A total of 226 individuals aged between 20 and 50 years participated in the listening test. The quality score ranges from 1-5, where a higher value indicates a better speech quality. The intelligibility score calculates the number of words correctly recognized by listeners in a ten-word sentence; the intelligibility score ranged from 0-10. A higher intelligibility score indicates that the listeners correctly identified more words. A total of 24,408 samples were collected. We followed the setup described in \cite{chen2021inqss} to divide the TMHINT-QI dataset into training and test sets. The subjective scores for each utterance were averaged to obtain its ground-truth score. The final training and test sets contained 12,937 and 1,978 unique utterances, respectively, along with their subjective quality and intelligibility scores. Further details are provided in \cite{chen2021inqss}.  

\subsection{Correlation analysis}
\label{sec:correllation}
We investigated the relationship between the subjective quality and intelligibility ratings and objective measures of the test data by calculating the Pearson correlation coefficient (PCC). The correlation values between the subjective and objective measures are presented in Fig. \ref{fig:Corr}, along with the correlation between subjective quality and intelligibility. Several observations can be drawn from the figures. Human perceptions of quality and intelligibility were moderately correlated, approximately 0.68. In addition, we observed that all objective measures except WER demonstrated higher correlations with subjective quality than with subjective intelligibility. 
\par For subjective quality, it is interesting to note that a high correlation is expected with PESQ, but objective intelligibility measures (i.e., NCM, ESTOI, STOI, and WER) are more highly correlated with subjective quality ratings. For subjective intelligibility, the correlations of objective quality measures (i.e., PESQ, ITU-T P.835, DNSMOS P.835, and MOSA-Net) were generally lower (below 0.24), which is reasonably expected because they were originally designed to predict speech quality. Interestingly, in relation to speech intelligibility, high correlations were expected between objective intelligibility measures (i.e., NCM, ESTOI, STOI, and WER); however, all measures, except WER, had moderate correlations with subjective intelligibility in our dataset. We also exploited the correlations between STOI, PESQ and WER. Fig. \ref{fig:DetailCorr} shows the scatter plots of the WER against the PESQ and STOI. Our findings are consistent with those of previous studies \cite{moore2017speech,siddiqui2020using}, which show that the correlation value between STOI and WER was higher than that between PESQ and WER. This supports the results in \cite{fu2018end} that integrating the STOI into the SE model optimization can improve the WER for enhanced speech. In summary, the strongest absolute correlation with subjective quality was found for the NCM, followed by the ESTOI and STOI. For subjective intelligibility, WER showed the highest absolute correlation, followed by subjective quality and NCM.  

\section{Experiments}
\label{sec:results}
\subsection{Experimental setup}
The correlation analysis in Section \ref{sec:correllation} indicates that none of the objective measures show a strong correlation (above 0.8) with subjective quality and intelligibility ratings, which aligns with the findings of \cite{hu2007evaluation} that no single objective measure demonstrates a high correlation. Thus, we sought to develop a DL model that uses a combination of objective measures as inputs to predict the corresponding subjective quality and intelligibility scores. 
\par
\begin{figure}[t!]
\begin{minipage}[b]{1\linewidth}
  \centering
  \centerline{\includegraphics[width=8.7cm]{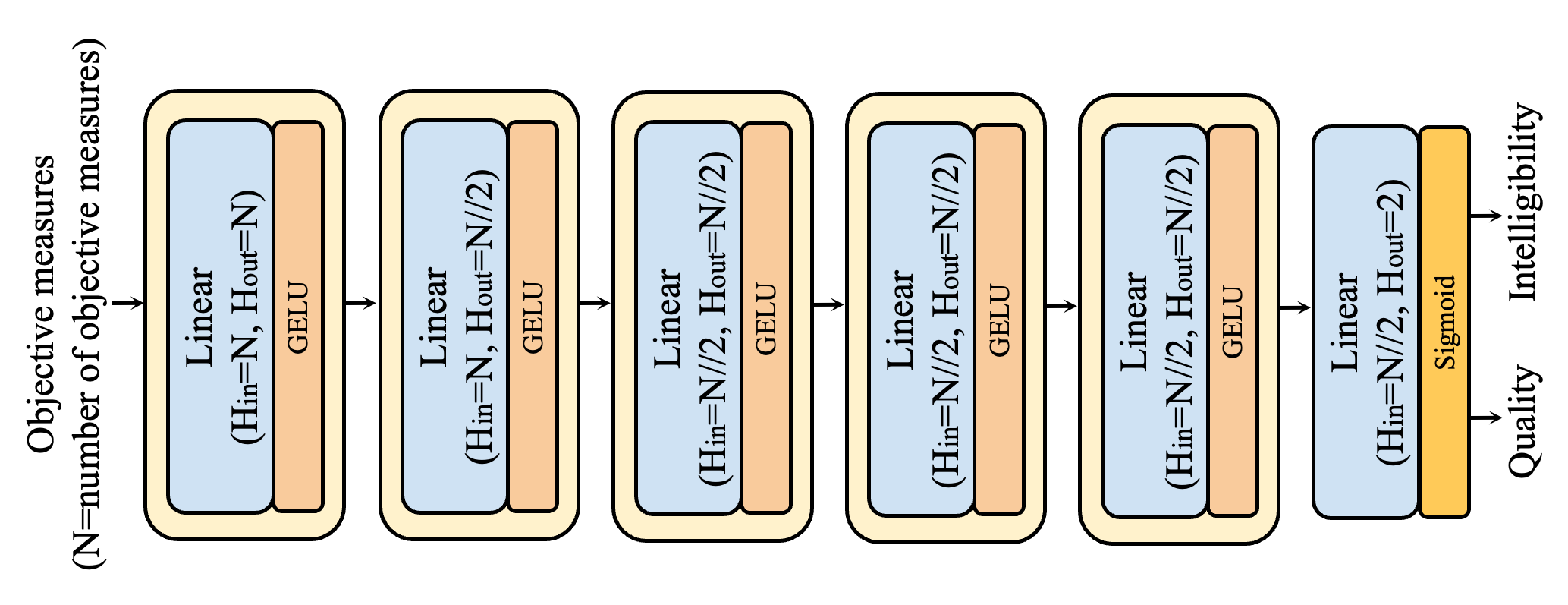}}
\end{minipage}
\caption{The proposed DL model receives twelve objective measures as input and generates outputs for human quality and intelligibility, where N equals twelve.}
\label{fig:model}
\end{figure}

 Fig. \ref{fig:model} illustrates the details of the proposed DL model. Each of the objective and subjective measures was normalized using a min–max to be between 0 and 1 before being fed into the DL model. Twelve objective measures were utilized as input for the DL model. The DL model consisted of six dense layers, each of which was followed by GELU activation, except for the last layer, which was followed by a sigmoid activation. This sigmoid activation produces values between 0 and 1, which are then divided into two separate tasks, one for quality estimation and the other for intelligibility prediction. Subsequently, the output values were denormalized to obtain the predicted subjective quality and intelligibility scores. To evaluate the performance, three criteria were selected: mean squared error (MSE), PCC, and Spearman’s rank correlation coefficient (SRCC).

\subsection{Experimental results}
We examined the performance of the proposed DL model in relation to the linear regression (LR) model, which separately predicts the subjective quality and intelligibility scores. In addition, we incorporated two DL-based non-intrusive speech assessment models for our evaluation. The first model, InQSS, combines self-supervised models with scattering transform and simultaneously predicts both subjective quality and intelligibility \cite{chen2021inqss}. The second model, MOS-SSL, utilizes fine-tuned features from wav2vec2.0 to predict MOS \cite{cooper2022generalization}. We trained the MOS-SSL model on the TMHINT-QI dataset using a single-task criterion to predict the quality and intelligibility scores as separate targets. 
\par The results have been summarized in both Table \ref{table:tableQI} and Table \ref{table:tableQI_nooverlap}. In Table \ref{table:tableQI}, we follow the setup described in \cite{chen2021inqss}, which randomly splits the training set into 90\% for training and 10\% for validation,  whereas in Table \ref{table:tableQI_nooverlap}, we are assessing the model's generalization by ensuring that there is no speaker overlap between the training and validation sets. From both tables, it is evident that the DL model consistently achieves superior PCC and SRCC scores in comparison to the LR model, InQSS, and MOS-SSL in most situations. These outcomes highlight the heightened accuracy and reliability of the DL model in predicting subjective quality and intelligibility.  

\begin{table}[t!]
    \caption{Quality and intelligibility prediction results using training and validation sets following the configuration described in \cite{chen2021inqss}. The term "system" is used to assess either speech quality (denoted as Q) or speech intelligibility (denoted as I), or both aspects simultaneously.}
    \centering
    \resizebox{1\linewidth}{!}{
    \begin{tabular}{lccccc}
    \toprule[0.4mm]
    &\multirow{2}{*}{System}& \multicolumn{2}{c}{Quality} & \multicolumn{2}{c}{Intelligibility}  \\
    \cline{3-6}
    && PCC$\uparrow$& SRCC$\uparrow$ & PCC$\uparrow$& SRCC$\uparrow$ \\
    \hline
        InQSS &  \multirow{2}{*}{Q+I} & \multirow{2}{*}{0.804} & \multirow{2}{*}{0.759} & \multirow{2}{*}{0.791} & \multirow{2}{*}{\textbf{0.730}} \\
    \cite{chen2021inqss} & & & & & \\
    
    \hdashline

    MOS-SSL & Q & 0.805 & 0.761  &-&- \\
    \cite{cooper2022generalization} & I & - & - & 0.774 & 0.67  \\
    \hline
    \multirow{2}{*}{LR} & Q & 0.797 & 0.751 &-&- \\
    & I & - & - & 0.739 & 0.676 \\
    \hdashline
    
    DL & Q+I & \textbf{0.806} & \textbf{0.763} &\textbf{0.797}&\textbf{0.730} \\

    \bottomrule[0.4mm]
    \end{tabular}%
    }
    \label{table:tableQI}
\end{table}

\begin{table}[t!]
    \caption{Quality and intelligibility prediction results to assess generalization, employing training and validation sets without any speaker overlap. The term "system" is used to assess either speech quality (denoted as Q) or speech intelligibility (denoted as I), or both aspects simultaneously.}
    \centering
    \resizebox{1\linewidth}{!}{
    \begin{tabular}{lccccc}
    \toprule[0.4mm]
    &\multirow{2}{*}{System}& \multicolumn{2}{c}{Quality} & \multicolumn{2}{c}{Intelligibility}  \\
    \cline{3-6}
    && PCC$\uparrow$& SRCC$\uparrow$ & PCC$\uparrow$& SRCC$\uparrow$ \\
    \hline
    \multirow{2}{*}{LR} & Q & \textbf{0.799} & 0.733 &-&- \\
    & I & - & - & 0.733 & 0.728 \\
    \hdashline
    DL & Q+I & 0.794 & \textbf{0.741} &\textbf{0.766}&\textbf{0.733} \\

    \bottomrule[0.4mm]
    \end{tabular}%
    }
    \label{table:tableQI_nooverlap}
\end{table}

\begin{table}[t]
    \caption{PCC and percentage change (denoted by PC) for different amounts of training utterances. The PC is calculated by dividing the decrease in PCC by the PCC obtained from 12,000 training utterances.}
    \centering
    \resizebox{1\linewidth}{!}{
    \begin{tabular}{lccccc}
    \toprule[0.4mm]

    \multirow{2}{*}{} & \multirow{2}{*}{Data\%} & \multicolumn{2}{c}{Quality} & \multicolumn{2}{c}{Intelligibility} \\
    \cline{3-6}
    && PCC$\uparrow$ & PC $\downarrow$ & PCC$\uparrow$ & PC $\downarrow$ \\
    \hline
    \multirow{4}{*}{InQSS \cite{chen2021inqss}} & 1.66\% & 0.236 & 70.35 & 0.262 & 66.54 \\
    & 5\% & 0.501 & 37.06 & 0.521 & 33.46 \\
    & 25\% & 0.771 & 3.14 & 0.723 & 7.66 \\
    & 100\% & 0.796 & - & 0.783 & -\\
    \hdashline
    \multirow{4}{*}{MOS-SSL \cite{cooper2022generalization}} & 1.66\% & 0.578 & 27.57 & 0.080 & 89.58 \\
    & 5\% & 0.675 & 15.41 & 0.407 & 47.01 \\
    & 25\% & 0.767 & 3.88 & 0.714 & 7.03 \\
    & 100\% & 0.798 & - & 0.768 & -\\
    \hdashline
    \multirow{4}{*}{DL} & 1.66\% & \textbf{0.752} & \textbf{6.47} & \textbf{0.688} & \textbf{12.91} \\
    & 5\% & \textbf{0.777} & \textbf{3.36} & \textbf{0.754} & \textbf{4.56} \\
    & 25\% & \textbf{0.796} & \textbf{1.00} & \textbf{0.786} & \textbf{0.51} \\
    & 100\% & \textbf{0.804} & - & \textbf{0.790} & -\\
    \bottomrule[0.4mm]
    \end{tabular}
    }
    \label{table:tableI}
\end{table}

\par
\begin{figure}[t!]
\begin{minipage}[b]{1\linewidth}
  \centering
  \centerline{\includegraphics[width=8.5cm]{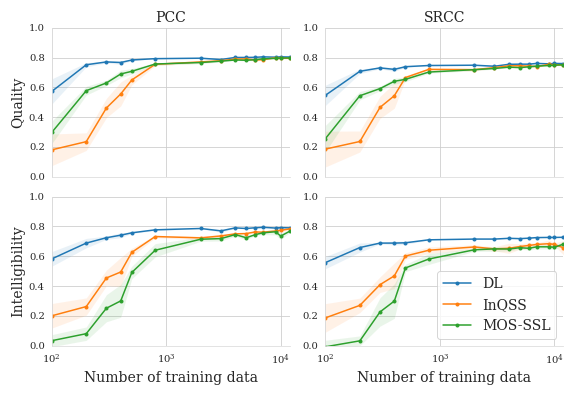}}
\end{minipage}
\caption{Correlations between prediction performance (PCC and SRCC) and training utterance quantity.}
\label{fig:diff_size}
\end{figure}

 We also examine the prediction accuracy of the proposed DL model compared with InQSS and MOS-SSL when different quantities of training data were accessible.  To avoid the time-consuming process of conducting listening tests, it is preferable to use a model that requires less training data but still provides comparable results. Table \ref{table:tableI} illustrates the percentage decrease in PCC for various percentages of training data, whereas Fig. \ref{fig:diff_size} visually represents the changes in the PCC and SRCC as the amount of training data varies. Table \ref{table:tableI} clearly shows that when trained with only 25\% of the data, all three models were close to reaching saturation. The InQSS and MOS-SSL models decreased by within 3\%, whereas the DL model for quality prediction decreased by 1\%. For intelligibility prediction, the InQSS and MOS-SSL models exhibited a decrease of 8\%, whereas the DL model exhibited a decrease of 5\%. Furthermore, the DL model demonstrated its superiority in performance with a percentage decrease of 3.4\% for quality prediction and 4.6\% for intelligibility prediction when trained using only 5\% of the training data. Fig. \ref{fig:diff_size} clearly demonstrates that the DL models consistently outperformed the InQSS and MOS-SSL models. Moreover, it shows that the increase in PCC and SRCC values gradually slows down when the amount of training data exceeds 1,000. The overall analysis indicates that InQSS and MOS-SSL rely heavily on a large amount of training data and exhibit promising prediction performances only when more training data are available. By contrast, the capacity of the proposed DL model to achieve good performance with a limited amounts of training data is a significant advantage. This is particularly beneficial because collecting subjective human ratings is a challenging and expensive process, and being less reliant on a large amount of data is highly advantageous.


\subsection{Interpretation of the DL model}
\begin{figure}[!t]
  \centering
  \centerline{\includegraphics[width=9.5cm]{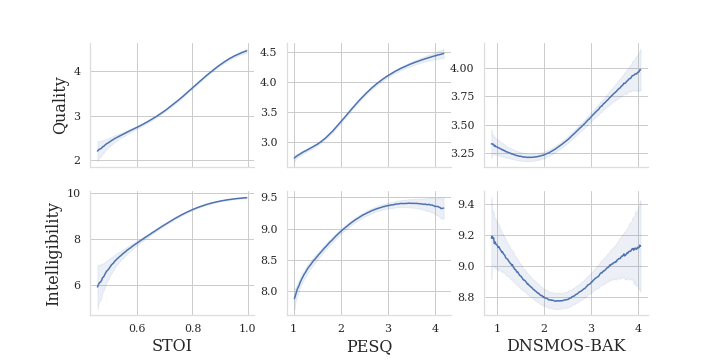}}
\caption{Plots illustrating the relationship between objective and subjective measures using the proposed DL model.}
\label{fig:funcrelationship}
\end{figure}

Our aim was to investigate how each objective measure affected the predictive performance of subjective quality and intelligibility. To uncover the underlying functional relationships between these measures in the DL model, we generated subjective quality and intelligibility scores by feeding data samples obtained from a multivariate normal distribution into the DL model. The subjective quality and intelligibility scores were then divided into 200 equal parts based on the values of the objective measures being analyzed. The scores for each part were averaged, resulting in 200 scores for each subjective quality and intelligibility. These scores were connected to form a line graph that illustrated the functional relationship between the quality or intelligibility scores and objective measures. This process was repeated 1,000 times and the functional relationship between the objective and subjective measures of the DL model is depicted in Fig. \ref{fig:funcrelationship}, where the solid line represents the mean, and the light-colored areas represent the standard deviation of the 1,000 lines. We limited our focus to several objective measures because of space limitations.

\par From Fig. \ref{fig:funcrelationship}, we can observe that the relationships between the objective measures and subjective quality appear to follow a fairly linear pattern, in contrast to subjective intelligibility. In the case of subjective intelligibility, it is apparent that the steepness of the slope gradually diminishes as the objective measures increase. This implies that higher values of objective measures can accurately demonstrate the expected improvement in subjective quality but not necessarily in subjective intelligibility. In addition, we found that the subjective measures declined when DNSMOS-BAK reached approximately 2.0. Specifically, there was a significant reduction in subjective intelligibility, decreasing from 9.2 to 8.8, while subjective quality experienced a less drastic drop, going from 3.4 to 3.2. Our findings suggest that this phenomenon occurs because attempts to suppress background noise inevitably result in speech distortion, which negatively affects the speech quality and intelligibility. Additionally, it is apparent that individual objective measures cannot fully capture subjective quality and intelligibility. This finding reinforces our rationale for integrating all objective measures to establish a strong correlation with subjective listening tests. 

\subsection{Enhancing intelligibility prediction through the incorporation of subjective quality}
Although our DL model predicts both subjective quality and intelligibility simultaneously, we are interested in exploring whether including subjective quality can enhance intelligibility prediction. The moderate correlation of 0.68 between subjective quality and subjective intelligibility indicates a potential association between the two factors. Consequently, we propose that integrating subjective quality ratings has the potential to enhance the prediction of subjective intelligibility, to some extent. Opting for quality tests instead of intelligibility tests provides significant advantages in terms of effort savings. Quality tests require less time compared to the time-consuming process of listening intelligibility tests, which involve word identification for calculating intelligibility scores. Therefore, the selection of quality tests is a more time-efficient approach. 
\par We modified the proposed DL model by including subjective quality scores as additional inputs. As a result, the model's primary objective shifted to predicting subjective intelligibility scores while considering these subjective quality scores. Table \ref{table:w/subq} shows a significant improvement in subjective intelligibility prediction when subjective quality ratings were incorporated. The PCC value increased from 0.792 to 0.870, validating the effectiveness of using subjective quality to predict intelligibility. The inclusion of subjective quality ratings represents a valuable contribution toward improving the accuracy of intelligibility predictions. By integrating subjective quality, we can harness the extensive research conducted in the field of speech quality assessment to enhance the assessment of speech intelligibility. 

\begin{table}[t!]
    \caption{Subjective intelligibility prediction results with objective measures alone (denoted by obj only) and a combination of objective and subjective quality measures (denoted by obj + sub Q). }
    \centering
    \begin{tabular}{lccc}
    \toprule[0.4mm]
    & MSE$\downarrow$ & PCC$\uparrow$ & SRCC$\uparrow$  \\
    \cline{1-3}
    \hline
    obj only & 1.771 & 0.793 & 0.726 \\
    \hdashline
    obj + sub Q & \textbf{1.234} & \textbf{0.870} & \textbf{0.756} \\
    \bottomrule[0.4mm]
    \end{tabular}
    \label{table:w/subq}
\end{table}

\section{Conclusion}
\label{sec:conclude}
The contributions of this study are as follows. First, this study proposes the use of DL models that use a combination of off-the-shelf objective measures as inputs to predict subjective quality and intelligibility ratings. Second, we evaluated the proposed DL model against different speech assessment methods and analyzed the percentage decrease in the PCCs as the amount of training data varied. The experimental results highlight the significant advantage of our DL model, which exhibits a strong performance even with a small amount of training data. This is particularly beneficial in situations in which gathering subjective human ratings is arduous and expensive. Thirdly, we provide insights into how objective measures reflect subjective quality and intelligibility. This analysis can help researchers better understand the relationship between objective and subjective measures. Fourthly, we demonstrated that incorporating subjective quality ratings can improve the prediction of subjective intelligibility. This integration allowed us to leverage the extensive research conducted in the field of speech quality assessment to enhance speech intelligibility evaluation. Additionally, quality tests offer a time-saving advantage over the more time-consuming process of listening intelligibility tests.



\bibliographystyle{IEEEbib}
\bibliography{strings}

\end{document}